\documentclass{llncs}
\pdfoutput=1
\usepackage[utf8]{inputenc}

\usepackage{graphicx}

\usepackage[hyphens]{url} 
\usepackage[breaklinks=true]{hyperref}
\usepackage{breakurl}

\usepackage{array}

\begin{document}

\author{Dominik Herrmann\inst{1} \and Jens Lindemann\inst{2}}

\title{Obtaining personal data and asking for erasure: Do app vendors and website owners honour your privacy rights?\thanks{This is the extended preprint of a paper published at ``Sicherheit 2016''.}}

\institute{
Universität Siegen, Institut f. Wirtschaftsinformatik, Kohlbettstr.~15, 57072 Siegen,\\ \email{herrmann@wiwi.uni-siegen.de}
\and
Universität Hamburg, FB Informatik, Vogt-Kölln-Str. 30, 22527 Hamburg,\\ \email{jens.lindemann@informatik.uni-hamburg.de}
}

\maketitle

\begin{abstract}
EU Directive 95/46/EC and the upcoming EU General Data Protection Regulation grant Europeans the right of access to data pertaining to them. Consumers can approach their service providers to obtain all personal data stored and processed there. Furthermore, they can demand erasure (or correction) of their data. We conducted an undercover field study to determine whether these rights can be exerted in practice. We assessed the behaviour of the vendors of 150 smartphone apps and 120 websites that are popular in Germany. Our deletion requests were fulfilled in 52 to 57\,\% of the cases and less than half of the data provision requests were answered satisfactorily. Further, we observed instances of carelessness: About 20\,\% of website owners would have disclosed our personal data to impostors. The results indicate that exerting privacy rights that have been introduced two decades ago is still a frustrating endeavour most of the time.
\end{abstract}


\section{Introduction}

Consumers are losing control of their personal data due to its pervasive collection in online services and apps.
Answering the simple question ``Who knows what about me?'', which is at the core of informational self-determination, has become a complex and involving task. However, users \emph{are} interested in answers. For instance, Lightbeam (previously called Collusion), which visualizes cookie-based tracking efforts, has been downloaded more than 600,000 times \cite{Collusion}. In a recent survey, more than 90\,\% of respondents stated that it was important for them (a) to be in control about \emph{who} gets information about them and (b) to control \emph{what} information is collected \cite{Pew15b}. Researchers have proposed \emph{privacy agents} for this purpose, client-side tools that keep track of the data entered into form fields \cite{KolterNP10}. However, privacy agents have not seen widespread adoption so far.

EU Directive 95/46/EC \cite{EUDirective} and the upcoming General Data Protection Regulation (GDPR) grant citizens the right to obtain all data relating to them from data controllers. Citizens can also demand rectification, erasure or blocking of their data. In Germany, these provisions have been implemented in the Federal Data Protection Act in 2009 (cf. \url{http://gesetze-im-internet.de/englisch_bdsg/}). They are mostly exercised for scoring (e.\,g. the German ``Schufa'') and marketing services (e.\,g. Arvato) at the moment. In principle, consumers can also exercise them for all online service providers, but their effectiveness is unknown so far. Will ordinary online shops bother to dig into their database when asked via e-mail by a customer?

Until now there was only circumstantial evidence suggesting that many data controllers are reluctant to answer data provision enquiries and that they fail to carry out deletion requests. Moreover, we wanted to evaluate whether data controllers handled personal data with due diligence. Therefore, we conducted two studies analysing the behaviour of vendors of services that are popular in Germany, comprising 150 smartphone apps and 120 websites. Both datasets have been compiled in the months of August and September of 2014.

This paper is structured as follows. First, we review related work in Sect.~2. In Sect.~3 we describe our methodology and present the results obtained for smartphone app vendors. Results of our experiments with website owners follow in Sect.~4. We discuss the results in Sect.~5, before we conclude the paper in Sect.~6.

\section{Related work}

We are not aware of previous research studying the behaviour of online service providers to determine the effectiveness of the right of access to personal data in practice. Balebako et al. \cite{Balebako14} analysed the privacy attitudes and the behaviour of app developers. However, in contrast to our undercover study, they conducted interviews and used an online survey.
Most research about online privacy focuses on software, e.\,g. app permissions as well as what kind of data is being collected by apps and whom it is shared with. Another line studies legal and usability aspects, for instance by analysing privacy policies. In the following we will review recent work along these lines.

Zang et al. \cite{Zang15} monitored HTTP(S) transmissions of popular apps and found that many apps share information with third-party websites. TaintDroid is an effort to detect privacy violations by means of taint analysis \cite{Enck10}.
The second GPEN Privacy Sweep \cite{PrivCA14} conducted by privacy enforcement authorities found that the majority of apps does not provide sufficient information for the user to understand why it is necessary to grant the requested permissions, which were found to be excessive in relation to the functionality for 31\,\% of the apps. Other studies are concerned with the willingness of smartphone users to grant apps certain permissions and/or their willingness to share personal information with third parties \cite{Lin12,Liu14}.
ICSI Haystack (\url{https://www.haystack.mobi}) alerts users about data leaks and collects data for research on privacy in mobile ecosystems.

Sunyaev et al. \cite{Sunyaev14} studied the 600 most popular mobile health apps and found that only about 30\,\% did have a privacy policy, out of which about two thirds ``did not specifically address the app itself''. They also deem these policies to be hard to read. Balebako et al. \cite{Balebako15} studied how well users will take notice of the privacy policy of an app depending on when it is being displayed. They found displaying the notice within the app to be more effective than displaying it in the app store. 
Schaub et al. \cite{Schaub15} analyse how privacy notices should be designed to be most effective, i.\,e. be noticed by users.
Vila et al. \cite{VilaGM04} state that users do not pay attention to privacy policies, as they are not a reliable indicator for whether the user's privacy will actually be respected by a website. A similar observation has been made by Chia et al. \cite{Chia12} for mobile apps: they found no reliable indicator for whether an app would actually respect a user's privacy.
Pollach \cite{Pollach07} also deems privacy policies to be ineffective in addressing user's privacy concerns due to them mostly being designed ``with the threat of privacy litigations in mind''. Through a linguistic analysis, she found that questionable uses of data are often downplayed in policies.
Despite users currently not paying attention to privacy policies, there is evidence that making the level of privacy easily visible to potential customers (e.\,g. by presenting it in a table) can influence their purchase decision, with them being willing to pay a premium for privacy \cite{Tsai11}.

So far, there is only little work in practice. The browser add-on ToS;DR (\url{https://tosdr.org}) enables users to better understand privacy policies. For selected sites, it displays a summarised version of the terms of service at the click of a button and provides a colour-coded rating. The website \emph{selbstauskunft.net} offers to send personal data enquiries to companies and authorities for German consumers, requiring virtually no effort. On \emph{justdelete.me}, consumers can find information on how to delete accounts with many online services.

\section{First study: popular smartphone apps}

\subsection{Data collection}

For the selection of popular apps we relied on AppAnnie (\url{http://www.appannie.com/}), a market research company that monitors the downloads of apps from Apple's and Google's stores. We downloaded AppAnnie's list of the 500 most popular apps (in terms of downloads in Germany). From this list we took the 25 most popular free as well as the 25 most popular paid apps each for Android and iOS (100 apps in total). In addition, we randomly selected 25 free apps per platform from apps ranked between 50 and 500 in order to extend the scope of the survey. If a specific app was available for both platforms, precedence was given to the Android version. The selected apps are listed in Appendix~1.

In the following we characterize the dataset (see Fig.~\ref{d1-desc} in the appendix). Entertainment apps and games dominate the dataset with a share of 46\,\% of all apps, followed by apps that are used for work and information exchange (39\,\%). The smallest groups are apps used for communication and shopping (8\,\% and 7\,\%, respectively). Most app vendors are located in Germany (38\,\%). Vendors from other European countries are responsible for 21\,\% of the apps in the dataset, while 36\,\% are from other parts of the world (mainly the US). For 5\,\% of the apps, it was not possible to determine the vendor's residence. While it is beyond the scope of this paper to profoundly assess the legal situation, we note that Section 1 of the German Federal Data Protection Act states that it \emph{is} applicable to vendors from non-EU countries under certain circumstances. While it may be difficult  to exercise privacy rights against vendors in foreign countries, we  still include those vendors in our study in order to compare their behaviour with that of domestic vendors.

We also tried to roughly categorize the vendors in terms of their size based on publicly available information, such as the number of employees. We found that 41\,\% of vendors are small, i.\,e. they consist of a single developer only, about a third are businesses with multiple developers, and 25\,\% are large enterprises.

\subsection{Methodology for interaction with app vendors}
\label{methodology-apps}

The first objective of the study is to assess how app vendors react when users contact them to request (a) provision of their personal data and (b) deletion of their personal data. In order to obtain realistic results, we did not want to disclose to them that we were conducting a study about their behaviour. Instead, we approached the vendors like ordinary consumers. The second objective is to validate the claims received by the vendors.

We downloaded each app from the app store onto a smartphone and used it for a couple of minutes. When an app allowed to create an account or asked to provide personal data, we entered as many details as possible. We made the following arrangements to avoid arousing suspicion. For each app, we set up a fake profile having a randomly chosen German name and a random birth date. When asked to choose login credentials, we derived the username from the name and year of birth, while the password was chosen randomly. For the postal address, we used that of the Computer Science Department of University of Hamburg, which provides a satisfactory cover because it also hosts a number of other companies. We instructed the campus post office to accept mail for our fake identities, even though it was not addressed to the Computer Science Department. Moreover, we registered the domain barmail.de and used e-mail addresses of the form \emph{firstname.lastname@barmail.de}.

We made a note of what data we entered in each app. Furthermore, we monitored the traffic of all apps to determine what pieces of data were disclosed to the vendors. For this purpose, we ensured that the apps were used while the smartphone was on a WiFi connection. We installed a custom CA root certificate on the phone to record all HTTP(S) traffic with BurpSuite (\url{http://portswigger.net/burp}).
When we were analysing the responses of the vendors at a later stage, the recorded traffic allowed us to \textbf{validate the claims} of the vendors regarding what pieces of information they have or have not received about our account and whether or not their apps were transferring sensitive pieces of data (e.\,g. IMSI/IMEI) to other parties.

One week later, we approached each vendor via e-mail. This \textbf{first data provision enquiry} consisted of an informal e-mail in German or English (depending on the language of the app). The mail was sent from the address used during account setup. If no account was set up in the app, the mail was sent from the e-mail address used to log in to the Apple iTunes or Google Play store. We sent the e-mail to the address stated on the app page within the respective store. If no address was stated there, we looked up a contact address on the vendor's website or submitted our enquiry via a web form provided there. In the enquiry we stated the respective e-mail address and username and asked the vendor to provide all personal data that was stored on its servers and the purpose of their collection. 

If there was no response within one week or if it was not satisfactory (cf. Sect.~\ref{sec:results1}), we sent a \textbf{second data provision enquiry}. The second enquiry was written more formally than the first one, because we hypothesized that a more formal and threatening tone would illicit more responses. Therefore, the enquiry contained an explicit reference to Section 34 of the German Federal Data Protection Act and in closing we announced that we would notify the supervisory authority according to Section 38 of the aforementioned act (which we did not in fact do) if we did not receive a response within seven days.

Four weeks after the last contact, we continued with the \textbf{analysis of account deletion practices}. This analysis was limited to those 56 apps allowing to set up an account. First, we tried to delete the account from within each app and on the vendor's website on our own.
Otherwise we sent a \textbf{first deletion request} to the vendor from the e-mail address used during registration, in which we informally asked to have our account deleted. If we did not receive a response within one week or if we determined that the account was not deleted completely, we followed up with a formal \textbf{second deletion request}, referring to Section 35 of the Data Protection Act. Again, we announced that we would notify the supervisory authority unless we received a response within seven days.
Obviously, it is impossible for consumers to verify that all of their personal data has been deleted completely by a vendor. Therefore, we can only look for signs that indicate an \emph{incomplete} deletion. We \textbf{verified claims about the purported deletion} of an account as follows: First, we tried to log in with the previously used credentials, which should not be possible any more. Second, we tried to reset the password, which, if successful, indicates that the account had been disabled rather than deleted. Third, we tried to create a new account with the same credentials. This should succeed. Otherwise, the vendor keeps some records about ``deleted'' accounts (which may in fact be compliant with the data protection act).

\subsection{Analysis and results for apps}
\label{sec:results1}

\begin{figure}[t]
\centering
 \includegraphics[width=1\columnwidth]{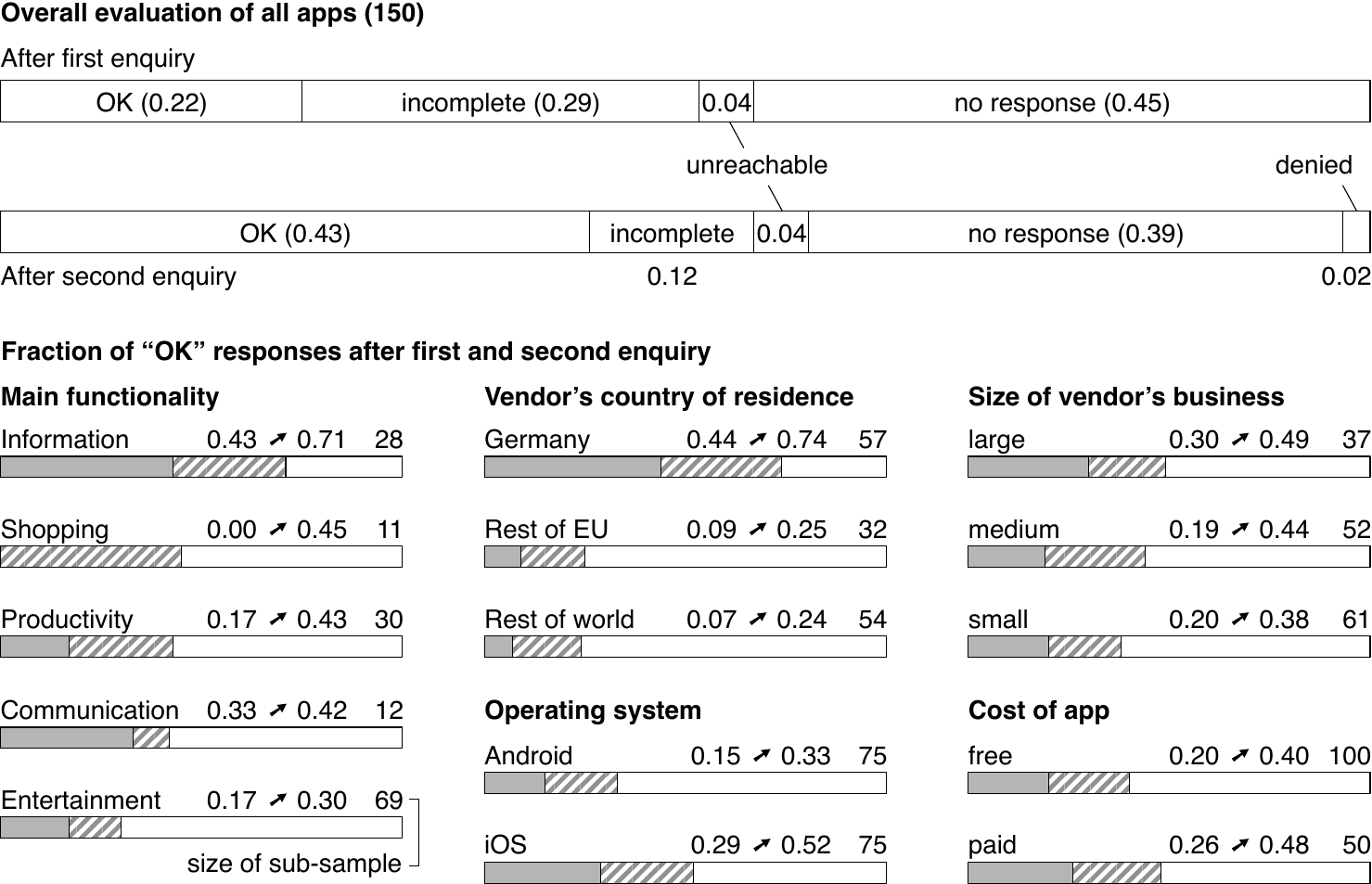}
 \caption{\label{r1}Results for data provision enquiry for popular apps}
\end{figure}

The results for the data provision enquiry are summarised in Fig.~\ref{r1}. The upper part shows the overall distribution of responses. A satisfactory \textbf{(OK)} response was scored if the response contained the actual data values that we had observed being transferred from the app to the vendor, or, in case we did not observe the app transmitting anything to the vendor at all, if the response stated this fact. The first enquiry was answered satisfactorily by only 22\,\% of vendors, increasing to 43\,\% after the more formal second enquiry. Two of the vendors responded by postal mail, all others via e-mail.
Some vendors only stated what kinds of information they typically store in an abstract fashion, i.\,e. they had not looked up our entry in their records at all. Others referred us to their privacy policy. Those cases were recorded as \textbf{incomplete} responses, because they did not provide our personal data as requested. The fraction of incomplete responses decreased from 29\,\% (first enquiry) to 12\,\% (second enquiry). This indicates that customers have to be insisting to obtain the desired information. For 4\,\% of the apps, we found no way to contact the vendor or our e-mail bounced \textbf{(unreachable)}. Many vendors did not reply to our mails (45 and 39\,\%, respectively). In response to the second enquiry, 2\,\% of vendors explicitly \textbf{denied} to answer our enquiry by e-mail, instructing us to call them or to send our enquiry by postal mail.

The bottom part of Fig.~\ref{r1} focuses on the ``OK'' cases. The highest fraction of satisfactory responses is achieved by German vendors (44\,\% and 74\,\%, respectively). One reason for this discrepancy may be that vendors from Germany are more sensitive to privacy than vendors from other countries. A reason for their poor performance after the second enquiry may be that \emph{foreign} vendors do not fear the consequences of customers notifying the German supervisory authorities, which cannot exert power in foreign jurisdictions.
Vendors of apps that are mainly used for providing or exchanging information scored better (71\,\%) than those of other apps (e.\,g. games and entertainment apps: 30\,\%). Further, vendors of iOS apps provided more satisfactory responses than those of Android apps (52 vs. 31\,\%, respectively). Vendors of paid apps scored only marginally better than those of free apps (48 vs. 40\,\%). Finally, large businesses scored best (49\,\%).

\begin{figure}[t]
\centering
 \includegraphics[width=1\columnwidth]{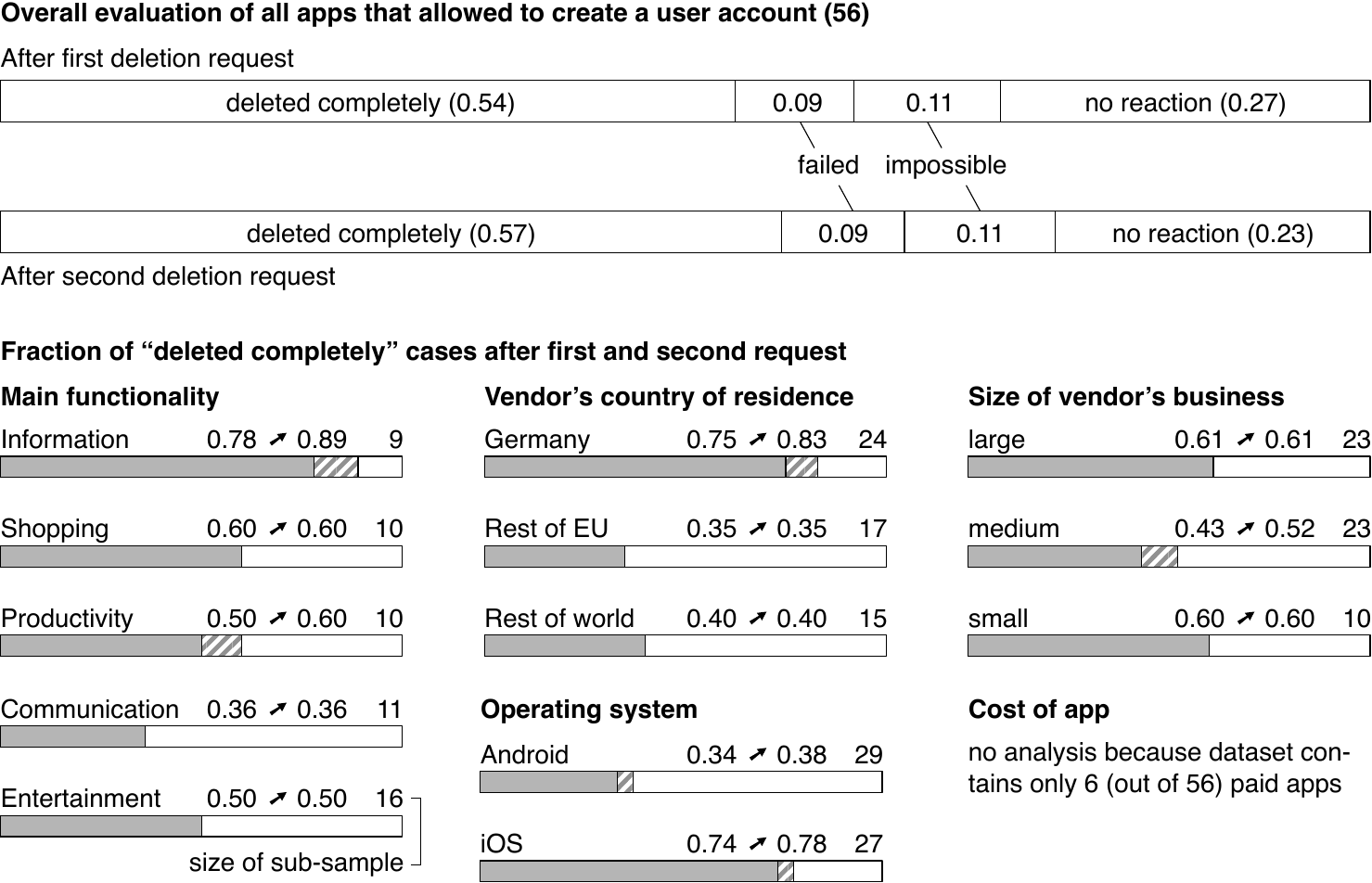}
 \caption{\label{r2}Results for deletion request for popular apps}
\end{figure}

Figure~\ref{r2} presents the results obtained for our deletion requests sent to the vendors of the subset of 56 apps in which we had created a user profile that was stored on the vendor's servers. We could delete 13 accounts on our own, either from within the app or on the vendor's website. Overall, deletion succeeded considerably more often than obtaining personal data. After the first informal request, 20 accounts were \textbf{completely deleted}. Out of the 13 accounts mentioned above, 10 were completely deleted, resulting in a total of 54\,\% of accounts (compared to 22\,\% satisfactory responses after the first data provision enquiry).
The second request increased this fraction only slightly to 57\,\%. All of our e-mails were delivered correctly this time, but in the end there was still \textbf{no reaction} in 23\,\% of all cases. In 5 cases (9\,\%; including 3 of the 13 accounts we were able to delete on our own), we were able to revive a ``deleted'' account \textbf{(deletion failed)}. The affected vendors declined to completely delete the accounts even after we confronted them with our finding (still using the fake profile as a cover). In 11\,\% of cases deletion was \textbf{impossible}: in half of these cases we couldn't find a way to contact the vendor, in the other cases support staff claimed that it was impossible to delete the account.

It may be perceived positively that all it takes for consumers to delete their account is writing an e-mail. Disturbingly, some vendors carried out deletion instantaneously, failing to check the identity of the sender. This lapse allows malicious users to delete the accounts of others by sending faked deletion requests with a forged sender address.

\section{Second study: popular websites}

\subsection{Data collection}

The dataset used in the second study contains popular websites that allow their users to create a user account to store personal data. The dataset was compiled based on the Alexa Top List for Germany (\url{http://www.alexa.com/}). First, we visited the 100 most popular websites. 57 of those 100 sites allowed us to set up an account and were therefore added to the dataset. Second, we randomly sampled 100 sites from the websites with ranks between 100 and 500. We added 63 of those sites to the dataset. Thus, the dataset contains 120 websites that are popular in Germany and allowed us to set up a user account (cf. Appendix~1). Some of the selected sites belong to vendors that have already been studied in the mobile app dataset. For these sites we duplicated the result obtained in the first study in order to avoid bias due to sending multiple queries to a vendor within a short period of time.
In contrast to the app dataset, shopping sites make up the largest part (40\,\%, compared to 7\,\% of apps), while the fraction of entertainment sites is quite small (12\,\%, compared to 46\,\% of apps). Further, there are much more large businesses in the website dataset (46\,\%, compared to 25\,\% of apps). Figure~\ref{d2-desc} in the appendix  shows the distributions in detail.

\subsection{Methodology for interaction with website owners}

The methodology used to study the behaviour of website owners resembles the procedure used for apps (cf. Sect.~\ref{methodology-apps}), i.\,e. we created a plausible fake user profile on each website, using e-mail addresses of the form \emph{firstname.lastname@barmail.de}. However, there is one important deviation: This time we also wanted to study how diligently site owners handled data provision enquiries. After all, the responses contain sensitive personal data. An impersonator should not be able to obtain personal data of a different person by fooling a website owner with a carefully crafted data provision enquiry. Spurious requests should be ignored and replies should only be sent to the e-mail or postal address stored in the user account.
For this analysis we registered the domain fair-konsult.de, which was to serve as a cover for a purported impostor. The impostor would send a data provision enquiry from the ``office address'' \emph{firstname.lastname@fair-konsult.de}. The text of the enquiry did not mention this discrepancy. Instead, it was an ordinary informal enquiry, i.\,e. it mentioned the \emph{firstname.lastname@barmail.de} address (and username, where appropriate) and stated the desire to obtain the personal data stored under that account.

If we received no satisfactory response from a site owner, we repeated the enquiry from the correct address, in analogy to the procedure used for apps. This time we created a second fake account on each site that was used for the deletion request. Having seen the results of the first study, we were concerned that our earlier provision enquiry might be stored in the customer support software of the site owner and bias the outcome of our deletion request.

\subsection{Results}

\begin{figure}[t]
\centering
 \includegraphics[width=1\columnwidth]{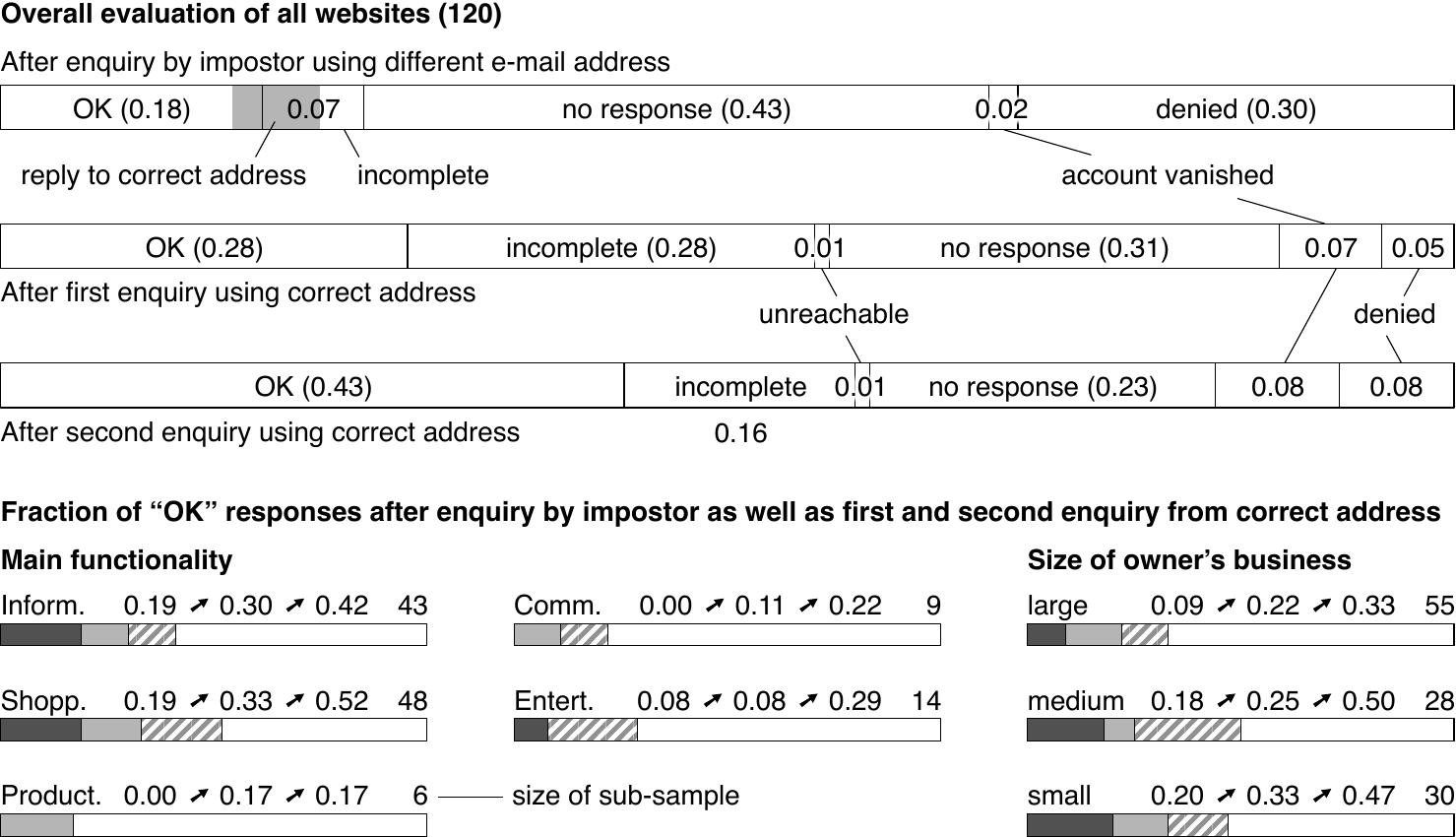}
 \caption{\label{r3}Results for data provision enquiry for popular websites}
\end{figure}

The results for the data provision enquiry are summarised in Fig.~\ref{r3}. 30\,\% of site owners noticed the incorrect e-mail address and responded with an explicit denial, while most (43\,\%) did not bother replying to our bogus enquiry at all. However, most interestingly, 25\,\% of website owners were indeed fooled into responding to our impostor with a satisfactory or incomplete response. One fifth of those replied to the \emph{barmail.de} address; the majority \emph{would} have disclosed sensitive personal data to an impostor, though. The breakdown of the results shows that this kind of fault happened at businesses of all sizes.
Including the ``OK'' cases obtained with the impostor's enquiry, 28\,\% of site owners have provided a complete and correct response to the first enquiry. Eventually, 43\,\% of responses were satisfactory. These and most of the remaining figures are in line with the results obtained for apps (cf. Sect.~\ref{sec:results1}). However, large businesses do not perform better this time. Moreover, some owners said they could not provide any data, because the account did not exist any more (8\,\% of accounts had \textbf{vanished} by the time of the second enquiry). In less than a handful of cases, site owners required the customer to call them to process the enquiry. As in the app study, only two site owners responded by postal mail.

\begin{figure}[t]
\centering
 \includegraphics[width=1\columnwidth]{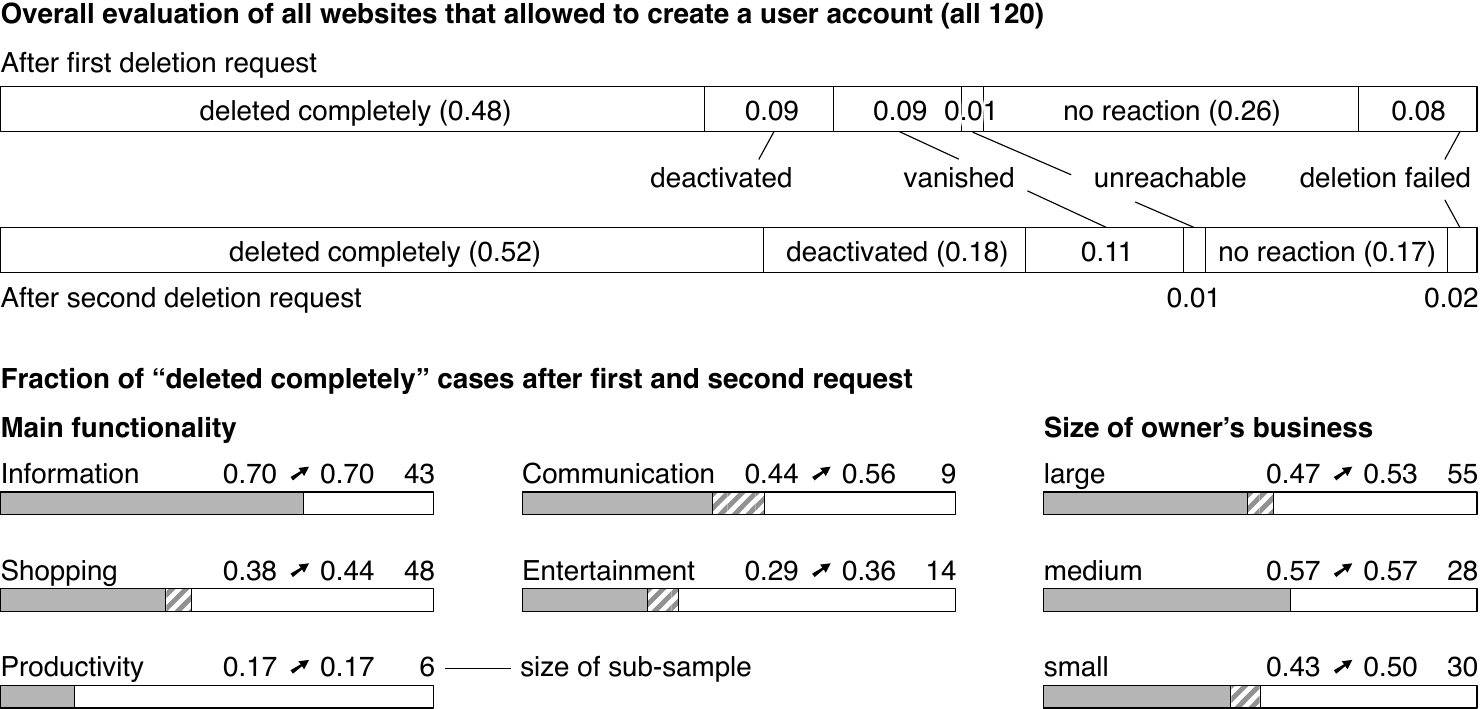}
 \caption{\label{r4}Results for deletion request for popular websites}
\end{figure}

Figure~\ref{r4} presents the results regarding the deletion requests. After the second request, 52\,\% of site owners had completely deleted our account. In contrast to the app study, some vendors (18\,\%) claimed that they could not delete the account for regulatory reasons (e.\,g. taxation). They promised to block the personal data. Again, a small fraction of site owners (11\,\%) claimed that there were no traces of the account in question (about eight weeks after it had been created). Apparently, some site owners purge inactive accounts after a while. This is in fact beneficial for privacy as it reduces the risk of data leaks for users.

\section{Discussion}

A typical ethical issue in undercover field research is that it consumes resources of the studied subjects without their consent, i.\,e. in our case the time of the employees of the surveyed service providers. However, we believe that this approach is the only way to investigate the problems related to data provision enquiries and deletion requests in practice. The insights gained may create awareness and sketch a path of action for regulatory authorities. In the design of our studies, we tried to minimize the burden on vendors.

While our objective was to survey popular apps and websites, we stress that the two samples were not meant to be representative. Therefore, the results cannot be generalized to the whole population of apps and websites. Further, the results of the first study may be biased due to giving preference to the Android version of an app in case it happened to be selected for both platforms. This means that apps that were very popular on both platforms have only been surveyed on Android, leading to the surveyed iOS apps being slightly less popular on average. Another potential problem is that it was classified as a failure to respond to our first enquiry if the response did not come within a week. As some responses arrived after this time frame had passed, it might be more appropriate to wait for a longer time in future studies. We are currently in the process of conducting a more extensive study, based on a more controlled and larger sample of smartphone apps.

Another potential source of bias stems from the design of the second study (websites), which includes an additional step to assess the viability of social engineering attacks. This additional query may have influenced the response of vendors. First, vendors that need considerably more time than we expected may have received our authentic mail (sent in the second step) \emph{before} they were about to reply to the faked mail. Second, vendors that have refrained from replying to our faked mail may actually have an increased inclination to respond once they receive our second mail (because it is the second contact for the same user account). Thus, these results are biased to some degree. In our follow-up study, we will therefore leave out the steps to analyse the viability of social engineering.

As noted in Sect.~\ref{methodology-apps}, it is impossible for us to verify whether a vendor has completely deleted personal information. This is why we have used techniques to infer whether any information is left behind. Another approach would have been to send another data provision enquiry after the deletion request. It would be interesting to see whether vendors would really admit that they have not complied with a deletion request. We leave this question for future work.

\section{Conclusion}

In this paper we quantified the effectiveness of the right of access to personal data with two studies. We found that for both apps and websites, only 43\,\% of our data provision requests were answered satisfactorily by vendors. Many of those who provided satisfactory information failed to do so after the first informal enquiry. Accordingly, obtaining one's personal data often requires perseverance and precise knowledge of the legal code. This may discourage many citizens from exercising this right. For deletion requests, the situation is somewhat better: Between 52\,\% to 57\,\% of accounts were deleted completely. This time, a large majority of vendors acted upon the first request, which may be due to the little effort involved. Worryingly, about 20\,\% of website owners were found to be susceptible to social engineering and provided the requested data to a different e-mail address than the one used to register the account. Although the right of access has been introduced about two decades ago, exercising it is both ineffective and inefficient. Instead of error-prone manual processing of such requests, service providers should (be legally bound to) offer consumers a standardised interface to access, rectify and erase/block their personal data.

\section*{Acknowledgements}

We are grateful to our students Jannik Schröder and Tim Richter. They interacted with vendors and site owners and provided preliminary analyses of the raw data.

\bibliographystyle{splncs03}
\bibliography{privacy-enquiry-arxiv}

\clearpage

\appendix

\section*{Appendix 1: Description of datasets}

\noindent The following apps were selected for the first study:

\medskip

\noindent\textbf{50 free Android apps:} Abs workout, ADAC Maps, Adobe Reader, Angry Birds Seasons,
Angry Birds Star Wars II Free, AutoScout24, Borussia Dortmund, Bubble
Island Adventure, Bubble Witch 2 Saga, Candy Crush Saga, Clash of
Clans, Clean Master Phone Boost, DB Navigator, Diamond Dash, Dolphin
Browser, eBay, eBay Kleinanzeigen, Facebook, Facebook Messenger, Fantasy
Warlord, Farm Heroes Saga, Flightradar24 Free, Focus Online, Head
Soccer, HVV, idealo, Instagram, Kritika, LINE Gratis-Anrufe, mehr-tanken,
Mobile Security \& Antivirus, Opera, Plants vs. Zombies 2, Plumber,
Shadow Kings, Shazam, Skype, Sonic Dash, SPORT1.fm, Spotify, Steam,
Subway Surfers, Swamp Attack, Toilet \& Bathroom Rush, TrackID, Traffic
Racer, Vector, Viber, wetter.com, Whats\-App

\medskip

\noindent\textbf{50 free iOS apps:} 1-2-3 Tanken,
Adobe Photoshop Express, BlaBlaCar, Brain+,
Buddyman:Office Kick, Bundesliga App Fußballfunk, Camu, Color Docks,
Cool Lock Screens, Dont Touch This!, Emoji for iOS 7, Fluege.de, Free
Music Mp3 Downloader for SoundCloud, fussball.de, Game of War Fire
Age, Gmail, Google Earth, Google Maps, Hay Day, KICK 15, King of Castles,
Kundencenter, Lieferheld, Lotto24.de, Lufthansa, Make Them Fall, McDonalds,
Mein wahres Ich, Mobile.de, MyVideo TV, Numbrs, Quick Scan QR Code
Reader, Retrica, RollerCoaster Tycoon 4 Mobile, Runtastic Six Pack,
Scuba Dupa, SIMSme, Sportschau, Star Wars: Commander, Stern Quiz Battle,
Swing Copters, Tango, The Rats Online, TV Movie, TV Smiles, Web.de
Mail, Wetter.de, Wikipedia, Wissenstraining Allgemeinbildung, Youtube

\medskip

\noindent\textbf{25 paid Android apps:} aCalendar+, Afterlight, Blitzer.de Plus, Construction Simulator
2014, Farming Simulator 14, Flightradar24 Pro, Gangstar Vegas, Geometry
Dash, GPS Navigation \& Maps, Hitman GO, Maps.me, Minecraft Pocket
Edition, Modern Combat 5 Blackout, Monopoly, Plants vs. Zombies, Poweramp
Full Version Unlock, Quizduell Premium, Runtastic PRO, Scribblenauts
Remix, Survival in Forest, TeamSpeak 3, The Game of Life, Threema,
True Skate, WeatherPro

\medskip

\noindent\textbf{25 paid iOS apps:} 7 Minute Workout, Akinator the Genie, All-in Fitness, Baby
Monitor 3G, Bodyweight, Crazy Chicken Deluxe, Doodle Jump, Facetune,
Fitness Point Pro, Freeletics, Geocaching, GoodReader, Maps 3D, miCal,
Micromon, Plague Inc., Pou, PowerCam, Quit Smoking Pro, R.TYPE, RegenRadar
Pro, Rules!, Star Walk 2, Tiny Wings, Weather Live Reloaded

\clearpage

\noindent The following websites were selected for the second study (the default top level domain ``de'' is not shown for reasons of conciseness):

\medskip

\noindent\textbf{57 websites with a rank between 1 and 100:}
adobe, amazon, apple, arbeits\-agentur, autoscout24, bahn, bild, booking.com, chefkoch, chip, computerbild, dict.cc, duden, ebay, facebook, faz.net, focus, fotolia, gameforge, gutefrage, handelsblatt, heise, immobilienscout24, kicker, linguee, linkedin, live.com, livejasmin, meinestadt, mobile, mydealz, paypal, pinterest, pornhub.com, redtube.com, rtl, spiegel, sport1, stackoverflow.com, stern, streamcloud, sueddeutsche, t3n, transfermarkt, tumblr.com, twitter.com, web, welt, wetter.com, wetter.de, xhamster.com, xing, xvideos.com, youporn.com, youtube, zalando, zeit

\medskip

\noindent\textbf{63 websites with a rank between 100 and 500:}
ab-in-den-urlaub, alternate, android-hilfe, asos, autobild, autotopdeal, billiger, bonprix, bwin, cam4, carstart, chattalk, comunio, conrad, cyberghostvpn, cyberport, dawanda, dropbox, evernote, expedia, expekt, finestautomotive, finya, flyeralarm, friendscout, fussball, github.com, hamburg, hm, homegate, ikea, immokat, immonet, immo\-welt, joyclub, kalaydo, klamm, klicktel, markt, mitfahrgelegenheit, mitfahrzentrale, monster, mybet, mytoys, notebooksbilliger, obi, otto, pcgameshardware, raumdirekt, saturn, soundcloud, spox, stayfriends, stellenanzeigen, stepstone, tchibo, thomann, tipico, twitch, urlaubspiraten, vierklee, vimeo, wikia

\begin{figure}[p]
\centering
 \includegraphics[width=1\columnwidth]{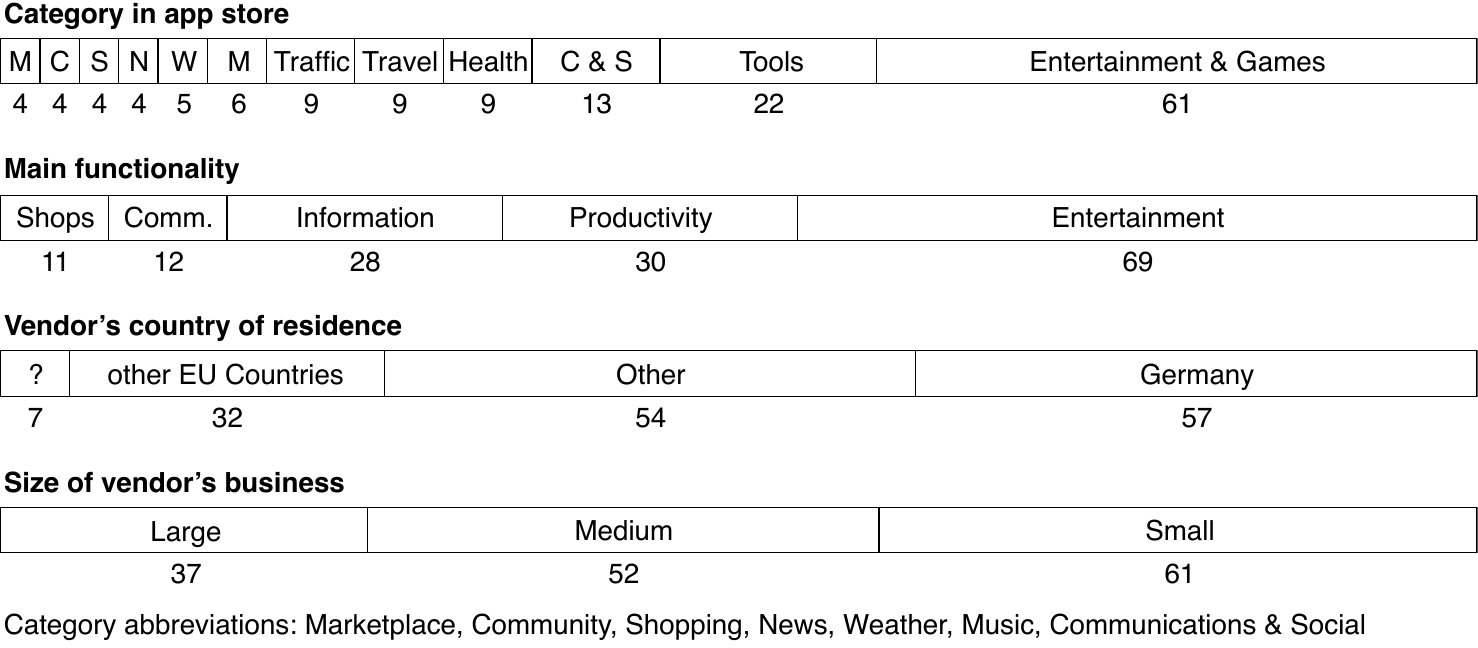}
 \caption{\label{d1-desc}Popular apps dataset overview; figures refer to number of apps out of 150 apps in total}
\end{figure}

\begin{figure}[p]
\centering
 \includegraphics[width=1\columnwidth]{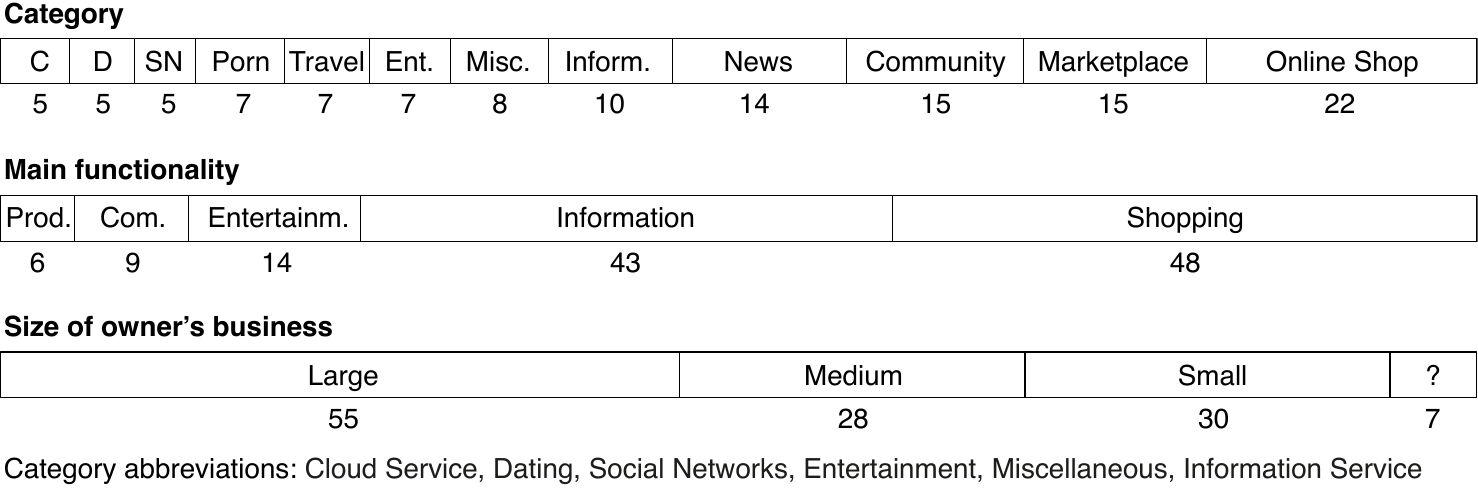}
 \caption{\label{d2-desc}Popular websites dataset overview; figures refer to number of sites out of 120 sites in total}
\end{figure}

\end{document}